\batchmode
\makeatletter
\def\input@path{{/home/yecheng/TCAS1-2018/}}
\makeatother
\documentclass[twocolumn,english]{IEEEtran}
\usepackage[T1]{fontenc}
\usepackage[latin9]{inputenc}
\usepackage{color}
\usepackage{array}
\usepackage{verbatim}
\usepackage{float}
\usepackage{textcomp}
\usepackage{multirow}
\usepackage{amsmath}
\usepackage{amssymb}
\usepackage{graphicx}

\makeatletter

\newcommand{\lyxmathsym}[1]{\ifmmode\begingroup\def\b@ld{bold}
  \text{\ifx\math@version\b@ld\bfseries\fi#1}\endgroup\else#1\fi}

\providecommand{\tabularnewline}{\\}
\floatstyle{ruled}
\newfloat{algorithm}{tbp}{loa}
\providecommand{\algorithmname}{Algorithm}
\floatname{algorithm}{\protect\algorithmname}

\usepackage{cite}

\usepackage{babel}
\usepackage{subfig}
\usepackage{caption}

\makeatother

\usepackage{babel}
\begin{document}
\title{\textcolor{black}{\pagenumbering{gobble}ChipNet: Real-Time LiDAR
Processing for Drivable Region Segmentation on an FPGA}}
\author{\textcolor{black}{Yecheng Lyu, Lin Bai and Xinming Huang~\IEEEmembership{Senior
Member,~IEEE}}\thanks{\textcolor{black}{This work is partially supported by by U.S. NSF
Grant CNS-1626236 and by The MathWorks fellowship. The authors are
with the Department of Electrical and Computer Engineering, Worcester
Polytechnic Institute, Worcester, Massachusetts 01609, USA. The corresponding
author is Xinming Huang (e-mail: xhuang@wpi.edu). }}\textcolor{black}{}\\
\textcolor{black}{{} }}
\maketitle
\begin{abstract}
\textcolor{black}{This paper presents a field-programmable gate array
(FPGA) design of a segmentation algorithm based on convolutional neural
network (CNN) that can process light detection and ranging (LiDAR)
data in real-time. For autonomous vehicles, drivable region segmentation
is an essential step that sets up the static constraints for planning
tasks. Traditional drivable region segmentation algorithms are mostly
developed on camera data, so their performance is susceptible to the
light conditions and the qualities of road markings. LiDAR sensors
can obtain the 3D geometry information of the vehicle surroundings
with high precision. However, it is a computational challenge to process
a large amount of LiDAR data in real-time. In this paper, a convolutional
neural network model is proposed and trained to perform semantic segmentation
using data from the LiDAR sensor. An efficient hardware architecture
is proposed and implemented on an FPGA that can process each LiDAR
scan in 17.59 ms, which is much faster than the previous works. Evaluated
using Ford and KITTI road detection benchmarks, the proposed solution
achieves both high accuracy in performance and real-time processing
in speed. }
\end{abstract}


\section{\textcolor{black}{Introduction}}

\textcolor{black}{In recent years, we have witnessed a strong increase
of research interests on autonomous vehicles\cite{Levinson_Stanford2011towards}\cite{Cosgun_Honda2017towards}.
Since the DAPRA Urban Challenge in 2007, automated driving technology
has grown rapidly from research experiments to commercial vehicle
prototypes owing to the explosive progress in the fields of artificial
intelligence and machine learning. As an important task of an automated
driving system, it is critical to conduct research on traffic scene
perception and its implementations on hardware platforms.}

\textcolor{black}{For traffic scene perception, detecting and tracking
algorithms are aimed to perceive the surroundings and to set the constraints
for planning and control tasks. Based on the object types, the task
of traffic scene perception can be classified into three sub-tasks:
(1) road perception includes drivable region segmentation and lane
detection, (2) object detection/tracking, and (3) traffic sign/signal
detection. In road perception, drivable region segmentation scans
the front area and searches for the drivable region, while lane detection
narrows the region of planning to the ego-lane if lane markers are
visible. Object detection and tracking identify the moving objects
such as vehicles, pedestrians, cyclists and animals, and measure their
locations, dimensions and speed to avoid a collision. Traffic sign/signal
detection looks for traffic signs and traffic lights to perceive additional
constraints for planning tasks \cite{chen2016accurate}. As a critical
component of an automated driving system, drivable region segmentation
provides fundamental knowledge of driving environment. Drivable region
segmentation solutions are required to perceive a wide range of view,
generate accurate results, and respond in real-time. However, road
scenes are complicated. As described in \cite{hillel2014recent},
road scenes have three types of diversities: (1) appearance diversity
due to changing shapes of lane markers and camera lens distortion,
(2) clarity diversity due to occlusions and illumination, and (3)
visibility condition diversity due to weather conditions. }

\textcolor{black}{Many sensing modalities have been used for drivable
region segmentation. Vision modalities \cite{long2015FCN}\cite{badrinarayanan2017segnet}\cite{garnett2017StixelNet2}
are frequently applied on drivable region segmentation for two major
reasons: (1) Vision modality is similar to human visual system and
most road markers have features in the visual domain, and (2) As a
passive sensor, visual camera provides high-resolution data with rich
features. By implementing multiple cameras, stereo vision \cite{chen20153d}
can provide depth information for drivable region segmentation. However,
due to the diversity in road scene, it is difficult to design a feature
descriptor that handles all visual cases and light conditions. In
addition, Shen et al. proposed a series of algorithms to cluster super-pixels
that could improve vision based semantic segmentation \cite{shen2017higher}
\cite{shen2014lazy}.}

\textcolor{black}{Light Detection And Ranging (LiDAR) is another major
modality often used by autonomous vehicles. By actively emitting laser
beams and measuring the 3D geometry around the vehicle using Time
of Flight (ToF), LiDAR can provide a few million geometric points
per frame with centimeter accuracy. In addition, LiDAR is not subjected
to environmental illumination. However, compared to vision modalities,
LiDAR points are sparse and do not contain any visual features employed
in traditional vision based algorithms. Several recent works studied
traffic scene perception involving LiDAR modality and proposed various
schemes for data arrangement, feature extraction and sensor fusion
with monocular vision. Much in depth studies are needed on LiDAR data
arrangement and feature extraction for accurate and efficient LiDAR
based drivable region segmentation.}

\textcolor{black}{In the past decades, drivable region segmentation
has been studied with different sensors and methodologies. A general
solution consists of four components: pre-processing, feature extraction,
detection and post-processing. Pre-processing includes noise removal,
data sampling and transformation. Feature extraction encodes local
features such as color, edge and texture from pre-processed data.
Detection applies manually defined or machine learning based models
to detect road area or lane boundaries. Lastly, post-processing suppresses
candidates to provide final results. }

\textcolor{black}{In traditional computer vision algorithms, those
four steps are totally separated and the extracted features are often
describable. However, manually defined features and detectors only
work well in normal conditions but cannot handle much variations on
the road. Machine learning especially convolutional neural network
(CNN) based algorithms combine feature extraction and detection together.
Pre-processed data are fed into a well-structured CNN with millions
of  parameters. Despite that features and detectors are hardly describable
visually, machine learning based road perception algorithms have significant
advantages in accuracy when compared with traditional computer vision
based approach. }

\textcolor{black}{For autonomous vehicles, both real-time processing
speed and low power consumption are desirable. Graphics processing
unit (GPU) devices are popular for parallel processing, but usually
consume too much power. Currently only one or two GPU devices can
be installed in a vehicle due to the limited power supply. But tens
of perception and planning tasks need to be processed on the GPUs
simultaneously. Field-programmable gate arrays (FPGAs) are low-power
devices that are more suitable for embedded systems. Moreover, an
FPGA can be developed as a customized integrated circuit that is able
to perform massive parallel processing and data communications on-chip.
Hereby, FPGA is our chosen platform that meets both computational
capability requirement and power consumption constraint. }

\textcolor{black}{In this paper, we present ChipNet as a CNN-based
algorithm and its FPGA implementation for real-time LiDAR data processing.
The contributions of our work can be summarized as follows: (1) We
introduce a new data organizing and sampling method in spherical coordinate
that improves the usage of LiDAR points and creates a dense input
tensor for CNN. (2) We propose an efficient convolution block for
CNN that is both hardware friendly and extendable. (3) The proposed
approach of drivable region segmentation results the state-of-art
accuracy when evaluated using Ford dataset and KITTI benchmark. We
also labelled the Ford dataset for training and evaluation. (4) An
efficient and flexible 3D convolution module is designed and implemented
on an FPGA, which can achieve real-time processing speed with limited
hardware resource and power usage.}

\textcolor{black}{The rest of the paper is organized as follows. Section
\ref{sec:Related-work} introduces the related works on road perception
task. The proposed drivable region segmentation algorithm is described
and its performance on benchmarks are presented in Section \ref{sec:Algorithm-design}.
Section \ref{sec:Hardware-Architecture} presents the FPGA architecture
and hardware implementation results. Finally, Section \ref{sec:Conclusions-and-future}
concludes the paper. }

\textcolor{black}{}
\begin{figure*}
\textcolor{black}{}%
\begin{tabular}{ccc}
\textcolor{black}{\includegraphics[height=3cm]{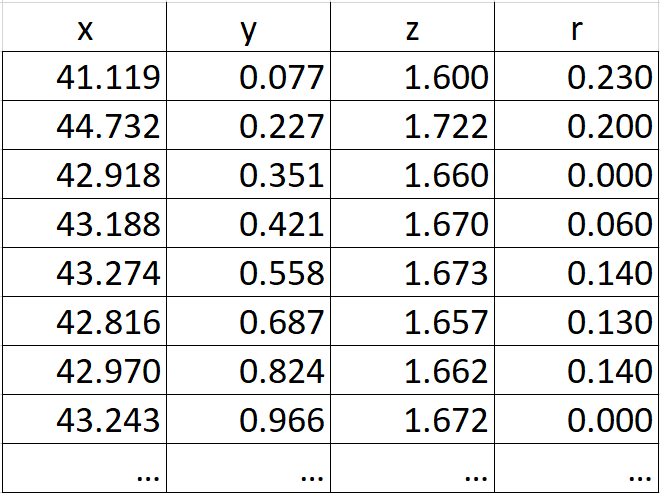}} &
\textcolor{black}{\includegraphics[height=3cm]{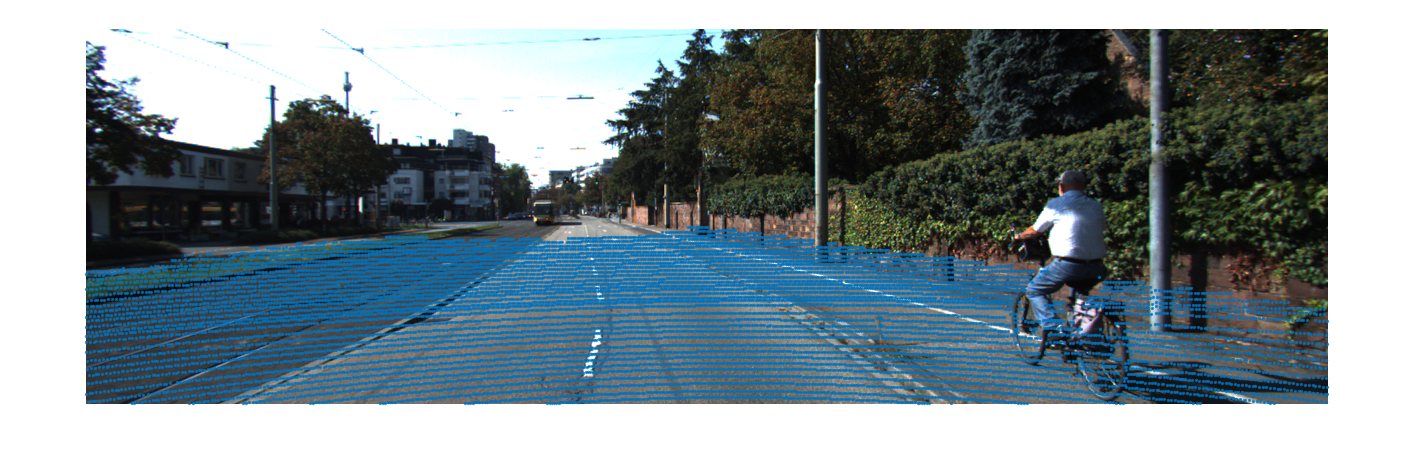}} &
\textcolor{black}{\includegraphics[height=3cm]{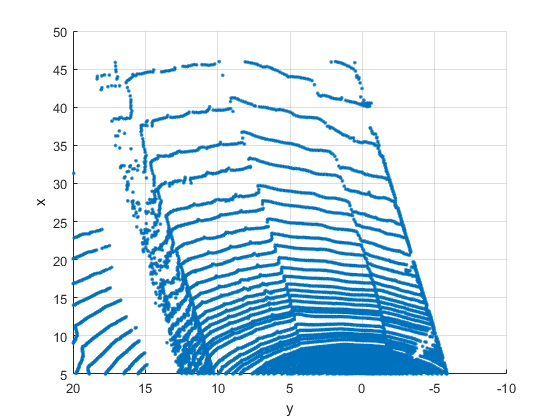}}\tabularnewline
\textcolor{black}{(a)} &
\textcolor{black}{(b)} &
\textcolor{black}{(c)}\tabularnewline
\end{tabular}

\textcolor{black}{\caption{A typical LiDAR frame. (a) LiDAR data matrix, (b) LiDAR points projected
on corresponding image, and (c) LiDAR points presented on top view
\label{fig:LiDAR-data-example}}
}
\end{figure*}

\section{\textcolor{black}{Related Work\label{sec:Related-work}}}

\textbf{\textcolor{black}{LiDAR data arrangement:}}\textcolor{black}{{}
There exists various methods of LiDAR data arrangement on traffic
scene perception. In Soquet et al. \cite{soquet2007road}, Alvarez
et al. \cite{alvarez2011road}, Shinzato et al. \cite{shinzato2014road}
and Liu et al. \cite{liu2015graph}, LiDAR point cloud was projected
to image view and manually defined features were applied based on
evaluation measurements with image patches. Similar image view was
employed by Han et al. \cite{Han_njust2017lidar_CRF} and Gu et al.
\cite{Gu_njust2017lidar_histogram} followed by feature extraction
using histogram. Gonzalez et al. \cite{gonzalez2015multiview} and
Xiao et al. \cite{Xiao2017HybridCRF} created a dense depth map from
point cloud and then combined the map with the camera data for their
machine learning based road boundary detector. Similarly, the multi-view
method \cite{chen_baidu2017multi_3d} transformed point cloud into
both image and top views and then combined with camera data for sensor
fusion using a CNN. In addition, VoxelNet\cite{Zhou_apple2017VoxelNetEL}
and 3D-FCN \cite{li_Trunk20173D_FCN} directly processed sparse LiDAR
data in world coordinate using convolutional neural network. LoDNN
\cite{caltagirone2017LoDNN} organized the point cloud into a top
view and then fed it into a CNN to generate a heat map representing
the possibility of drivable region in each $0.1m\times0.1m$ cell. }

\textcolor{black}{Beside road perception, several research works proposed
using CNN for LiDAR-based vehicle detection \cite{Zhou_apple2017VoxelNetEL}\cite{chen_baidu2017multi_3d}.
To overcome the shortage of training samples, data augmentation and
coarse labeling methods were proposed to enlarge the dataset. VoxelNet
\cite{Zhou_apple2017VoxelNetEL} augmented training data by rotating
and translating LiDAR points together with ground truth. StixelNet
\cite{garnett2017StixelNet2} used LiDAR points to generate coarse
labeling automatically for pre-training.}

\textbf{\textcolor{black}{CNN for road perception:}}\textcolor{black}{{}
Convolutional neural networks have become an active approach for the
task of road perception. Starting from Fully Convolutional Network
(FCN) \cite{long2015FCN}, various network structures have been proposed
to provide accurate road detection and segmentation. SegNet \cite{badrinarayanan2017segnet}
introduced an encoder-decoder scheme to separate feature extractor
and detector components. It also added additional connections between
the encoder and decoder layers that improved the training of the first
few layers closer to the input. Oliveira et al. \cite{oliveira2016up-conv}
followed the encoder-decoder scheme and perceived near range and far
range in separate branches that resulted an increased accuracy of
vision based segmentation. RBNet \cite{chen2017rbnet} also followed
the encoder-decoder scheme but connected all encoder layer outputs
to the decoders. Other works introduced the CNN for salient object
detection in images \cite{wang2018deep} and videos \cite{wang2018video}.
Most recently, CNN has also been introduced to LiDAR based road segmentation.
LoDNN \cite{caltagirone2017LoDNN}, VoxelNet \cite{Zhou_apple2017VoxelNetEL}
and Multi-view \cite{chen_baidu2017multi_3d} proposed different techniques
on LiDAR based perception.}

\textbf{\textcolor{black}{Embedded platforms for road perception:}}\textcolor{black}{{}
Considering the situations of automated driving or advanced driver
assistance system (ADAS), the processing time of road perception algorithms
must fulfill the real-time requirement, and thus are often implemented
on embedded platforms such as FPGA, ASIC or a mobile CPU/GPU processor.
Huval et al. \cite{huval2015empirical} deployed a neural network
on Jetson TK1 mobile GPU platform. It detected lane markers based
on images and achieved 2.5 Hz running speed. Similarly, the neural
network proposed in \cite{wang2013fpga} was able to segment multiple
objects including vehicles, pedestrian and pavements at 10 Hz with
image resolution of 320p on TX1 GPU platform. Two FPGA based lane
detection solutions were proposed in \cite{zhao2014real} \cite{wang2013fpga}
and their processing time were at 60 Hz and 550 Hz, respectively.}

\section{\textcolor{black}{Algorithm Design\label{sec:Algorithm-design}}}

\textcolor{black}{In this work, a hardware friendly and extendable
convolutional neural network is proposed to segment drivable region
using LiDAR data. In this section, we first introduce the LiDAR data
preparation method as pre-processing of the CNN. Next, the proposed
network architecture ChipNet is described in detail. Furthermore,
we introduce a simulated quantization scheme for CNN that transforms
floating-point to fixed-point operations, and thus speeds up the processing
on hardware considerably. Finally, a post-processing algorithm is
developed to generate a decision map denoting the drivable regions
from the CNN output. The proposed solution is evaluated on Ford Campus
Vision and LiDAR dataset and KITTI road benchmark. The performance
results are presented towards the end of this section.}

\textcolor{black}{}
\begin{figure*}[tp]
\begin{centering}
\textcolor{black}{\includegraphics[width=0.85\textwidth]{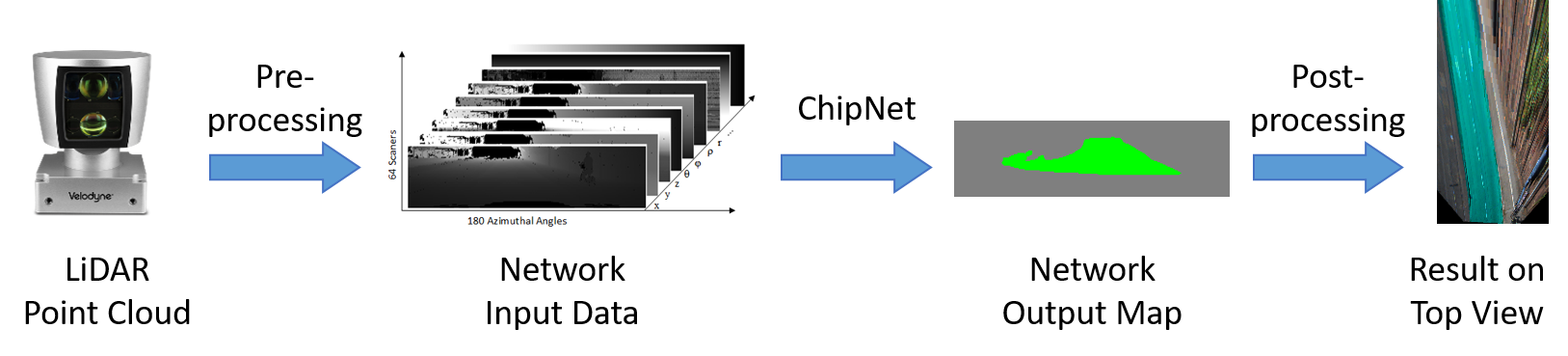}}
\par\end{centering}
\centering{}\textcolor{black}{\caption{Data Flow of the proposed LiDAR processing approach}
}
\end{figure*}

\subsection{\textcolor{black}{LiDAR data preparation\label{subsec:LiDAR-data-preparation}}}

\textcolor{black}{Typically a LiDAR device places a number of laser
scanners vertically and rotates them azimuthally to scan the surrounding
obstacles. Suppose a LiDAR device that contains $N$ scanners, measures
$M$ points per second and rotates at $R$ rpm, then it generates
$\frac{R}{60}$ frames per second with $\frac{60M}{R}$ measure points
per frame at an azimuthal resolution of $\frac{360NR}{60M}$. The
polar resolution is $\frac{\phi}{N}$ where $\phi$ denotes the vertical
field of view. For example, the HDL-64E LiDAR used in KITTI road benchmark\cite{Fritsch_KIT2013KITTI_road}
has $64$ scan channels and emits $1.33$ million points per second.
By rotating at $600$ rpm it updates $10$ frames per second with
$0.133$ million measurement points per frame at $0.17\lyxmathsym{\textdegree}$
azimuthal resolution. By focusing on a $26.90\lyxmathsym{\textdegree}$
vertical field of view, the polar resolution is $0.42\lyxmathsym{\textdegree}$.
In practice, the LiDAR sensor occasionally generates void measure
points when the laser beam emits to a low reflective surface.}

\textcolor{black}{Typically, a frame of data generated by the LiDAR
modality is a table as shown in Figure \ref{fig:LiDAR-data-example}.
In each row, the measurement of a corresponding LiDAR point is listed
in four columns, including location coordinates $x,y,z$ of the LiDAR
view and laser reflection intensity of the target surface $r$. By
projecting all points to camera view and top-view, as presented in
Figure \ref{fig:LiDAR-data-example}, LiDAR point cloud is sparse
and has large variations of point density throughout the entire space.
Therefore, LiDAR data needs to be organized and re-sampled before
being fed to the convolutional neural network. }

\textcolor{black}{As mentioned in Section \ref{sec:Related-work},
there is no unified method to LiDAR point cloud data arrangement and
sampling view. Table \ref{tab:Pre-processing} summarizes several
research works that organized LiDAR data in different forms. In Table
\ref{tab:Pre-processing}, we can see that most of them divided the
3D space in Cartesian coordinates, but they sampled the point cloud
in different views, such as top view \cite{caltagirone2017LoDNN}\cite{chen_baidu2017multi_3d},
front view \cite{Zhou_apple2017VoxelNetEL} or 3D view \cite{li_Trunk20173D_FCN}.
We also find a large percentage of LiDAR points are encoded in their
region of interest (RoI). However, the organized data as the input
to neural networks are sparse, which means that the majority of computations
in the first few layers of CNN actually deal with zeros. That is very
inefficient from the computational perspective.}

\begin{figure}
\begin{centering}
\textcolor{black}{\includegraphics[width=0.65\columnwidth]{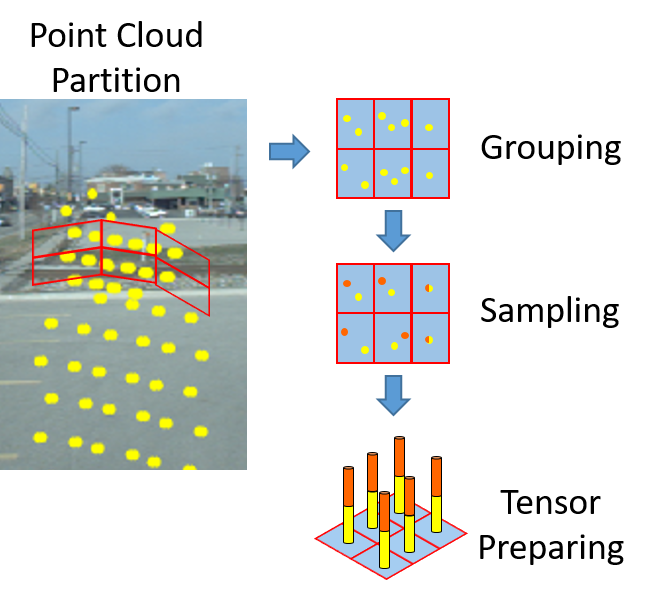}}
\par\end{centering}
\textcolor{black}{\caption{An illustration of the proposed LiDAR data preparation method\label{fig:LiDAR-data-preparation}}
}

\end{figure}

\textcolor{black}{Therefore, we propose to organize the LiDAR data
in spherical view as if a LiDAR naturally scans the surroundings,
as shown in Figure \ref{fig:LiDAR-data-preparation}. A region of
interest is selected in azimuth $[-45\lyxmathsym{\textdegree},45\lyxmathsym{\textdegree})$
and all $64$ lines of scan points are involved in segmentation. On
each line, scan points are grouped by every $0.5\lyxmathsym{\textdegree}$
into cells. In total, all scan points in the RoI fall into a $180\times64$
mesh. We use $0.5\lyxmathsym{\textdegree}$ because it is $3$ times
of LiDAR azimuthal resolution so that in theory at least $2$ scan
points are grouped in each cell. In practical terms, there are some
void scans when the reflect surface is out of range or has low-reflectivity.
Input tensor is built in the same width and height as the scan point
mesh, but contains $14$ feature channels. In each cell, the first
$7$ features come from the point nearest to the scanner, the next
$7$ features come from the point furthest away from the scanner.
These features include Cartesian coordinates $x,y,z$, spherical coordinates
$\theta,\varphi,\rho$ and the laser reflection intensity $r$. }

\textcolor{black}{In Table \ref{tab:Pre-processing}, we compare the
LiDAR data preparation with several related works. By sampling LiDAR
points in spherical view, our work not only has high LiDAR point usage
in RoI but also creates a dense input tensor that improves the accuracy
performance and makes the computations in the CNN much more efficient.}

\textcolor{black}{}
\begin{table*}
\centering{}\textcolor{black}{\caption{Summary of LiDAR data organization as input to neural networks \label{tab:Pre-processing} }
}%
\begin{tabular}{|c|c|c|c|c|}
\hline 
\textcolor{black}{Method} &
\textcolor{black}{Space subdivision} &
\textcolor{black}{Sampling view} &
\textcolor{black}{LiDAR point usage in input tensor} &
\textcolor{black}{Cell usage in input tensor}\tabularnewline
\hline 
\textcolor{black}{VoxelNet\cite{Zhou_apple2017VoxelNetEL}} &
\textcolor{black}{even on Cartesian} &
\textcolor{black}{front view} &
\textcolor{black}{98\%} &
\textcolor{black}{1\%}\tabularnewline
\hline 
\textcolor{black}{3D-FCN\cite{li_Trunk20173D_FCN}} &
\textcolor{black}{even on Cartesian} &
\textcolor{black}{3D view} &
\textcolor{black}{32\%} &
\textcolor{black}{1\%}\tabularnewline
\hline 
\textcolor{black}{Multi-view (top-view)\cite{chen_baidu2017multi_3d}} &
\textcolor{black}{even on Cartesian } &
\textcolor{black}{bird eye view } &
\textcolor{black}{21\%} &
\textcolor{black}{15\%}\tabularnewline
\hline 
\textcolor{black}{Multi-view (front-view)\cite{chen_baidu2017multi_3d}} &
\textcolor{black}{uneven on camera view} &
\textcolor{black}{camera view} &
\textcolor{black}{73\%} &
\textcolor{black}{8\%}\tabularnewline
\hline 
\textcolor{black}{LoDNN\cite{caltagirone2017LoDNN}} &
\textcolor{black}{even on Cartesian} &
\textcolor{black}{bird eye view} &
\textcolor{black}{24\%} &
\textcolor{black}{15\%}\tabularnewline
\hline 
\textcolor{black}{ChipNet (ours)} &
\textcolor{black}{even on Spherical} &
\textcolor{black}{spherical view} &
\textcolor{black}{66\%} &
\textcolor{black}{87\%}\tabularnewline
\hline 
\end{tabular}
\end{table*}
\textcolor{black}{}

\subsection{\textcolor{black}{ChipNet: a hardware friendly and extendable CNN
architecture}}

\textcolor{black}{In this section, we introduce the ChipNet architecture
and its simulated quantization algorithm when training on a GPU. The
innovations of ChipNet are: (1) we designed a dilated block equivalent
to a $5\times5$ convolutional kernel but saves parameters and calculations,
(2) we designed an extendable CNN structure using the dilated block,
and (3) we proposed a simulated quantization algorithm to obtain the
fixed-point parameters for hardware implementations. The network is
evaluated using Ford dataset and KITTI road benchmark.}

\subsubsection{\textcolor{black}{ChipNet convolutional block}}

\begin{figure}[h]
\begin{centering}
\textcolor{black}{}%
\begin{tabular}{cc}
\textcolor{black}{\includegraphics[height=4cm]{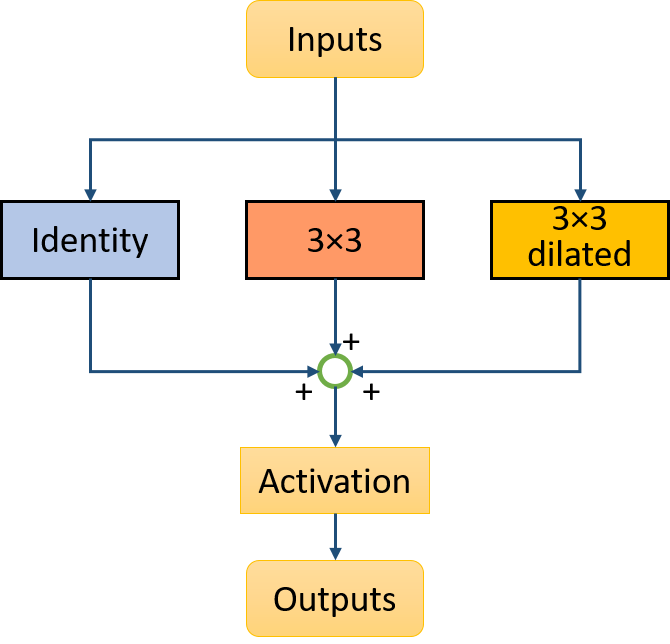}} &
\textcolor{black}{\includegraphics[height=4cm]{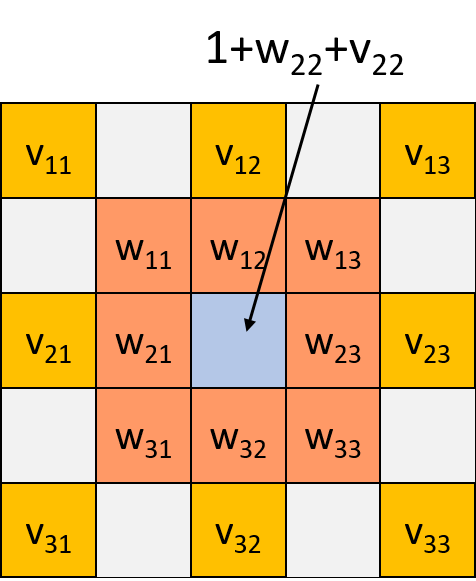}}\tabularnewline
\textcolor{black}{(a)} &
\textcolor{black}{(b)}\tabularnewline
\end{tabular}
\par\end{centering}
\textcolor{black}{\caption{ChipNet convolution block. (a) block architecture and (b) its equivalent
$5\times5$ convolution kernel. The cells in blue, red and brown denote
the contributions of corresponding convolutional operations in a conventional
$5\times5$ convolution kernel.\label{fig:Block}}
}
\end{figure}

\textcolor{black}{The convolutional block is a key component in ChipNet
architecture. Each network block contains three branches. The first
one is an identity branch that directly copies the input to the output.
As analyzed in \cite{he_FAIR2016resnet}, identity branch contributes
the majority of gradient in back-propagation and decreases the chance
of gradient vanishing and explosion during training. The second branch
is a $3\times3$ convolutional layer with $64$ channel outputs. The
second branch is aimed to encode local features. The third branch
is a dilated $3\times3$ convolutional layer \cite{Yu2016dilated_conv}
to process features in further pixels but takes less parameters and
calculations. As shown in Figure \ref{fig:Block}, after adding all
three branches element-wise, the block equivalents to a $5\times5$
convolutional layer but has a stable gradient in back-propagation
and fewer parameters. Assuming the convolutional layer input size
is $180\times64\times64$ and output size is also $180\times64\times64$,
a ChipNet block contains only 73,856 parameters and requires 802 million
multiplications. In comparison, a conventional $5\times5$ convolutional
layer employs 102,464 parameters and requires 1,180 million multiplications.
As a results, the proposed ChipNet block reduce parameters by 28\%
and multiplications by 32\%.}

\textcolor{black}{}

\subsubsection{\textcolor{black}{ChipNet network architecture}}

\textcolor{black}{The overall CNN architecture of ChipNet is shown
in Table \ref{tab:ChipNet-architecture}. The first layer is a local
feature encoder aimed to encode the input LiDAR data into a 64-channel
feature tensor. After encoding, the proposed ChipNet convolution block
is instantiated repetitively in the network to perform additional
encoding and decoding. Since the input and output of all ChipNet blocks
are exactly in the same sizes, the neural network can be conveniently
extended deeper by adding more layers. In our work, the ChipNet block
is instantiated 10 times as a trade-off between segmentation accuracy
and processing latency. For the output layer, a channel-wise mapping
is used to generate the final decision map showing the probability
of corresponding drivable regions. Compared to FCN and SegNet, the
proposed network is much simpler and more importantly it is extendable.
The repetitive network structure is best fitted for hardware reuse
in the FPGA design. }

\textcolor{black}{}
\begin{table}
\textcolor{black}{\caption{Layer configuration of the ChipNet architecture\label{tab:ChipNet-architecture}}
}
\centering{}\textcolor{black}{}%
\begin{tabular}{|>{\centering}m{0.1\columnwidth}|>{\centering}m{0.25\columnwidth}|>{\centering}m{0.2\columnwidth}|>{\centering}m{0.2\columnwidth}|}
\hline 
\textcolor{black}{Layer} &
\textcolor{black}{Kernel}

\textcolor{black}{$(w\times h\times m\times n)$} &
\textcolor{black}{Input}

\textcolor{black}{$(w\times h\times m)$} &
\textcolor{black}{Output}

\textcolor{black}{$(w\times h\times n)$}\tabularnewline
\hline 
\textcolor{black}{Input} &
\textcolor{black}{$-$} &
\textcolor{black}{$180\times64\times14$} &
\textcolor{black}{$-$}\tabularnewline
\hline 
\textcolor{black}{Conv}

\textcolor{black}{Encoder} &
\textcolor{black}{$5\times5\times14\times64$} &
\textcolor{black}{$180\times64\times14$} &
\textcolor{black}{$180\times64\times64$}\tabularnewline
\hline 
\textcolor{black}{ChipNet}

\textcolor{black}{Block}

\textcolor{black}{$\times10$} &
\textcolor{black}{$\begin{array}{c}
1\times1\times64\times64\\
3\times3\times64\times64\\
3\times3\times64\times64
\end{array}$} &
\textcolor{black}{$180\times64\times64$} &
\textcolor{black}{$180\times64\times64$}\tabularnewline
\hline 
\textcolor{black}{Output } &
\textcolor{black}{$1\times1\times64\times1$} &
\textcolor{black}{$180\times64\times64$} &
\textcolor{black}{$180\times64\times1$}\tabularnewline
\hline 
\end{tabular}
\end{table}

\subsubsection{\textcolor{black}{Simulated quantization}}

\textcolor{black}{Simulated quantization is essential to the training
of networks on hardware. Fixed-point variables, weights and operations
are widely used in FPGA design, which often utilizes less hardware
resources and memories and results higher clock speed, if compared
with the floating-point implementations. However, CPU and GPU platforms
generally employ floating-point operations that have no quantization
error and can generate continuous gradients in the training session.
Practically, we can implement CNNs on an FPGA for low-power embedded
application. But we still heavily rely on the high-performance GPUs
to train the neural networks in order to generate the parameters and
weights, since a GPU machine is capable of storing terabytes of training
samples and processing hundreds of threads simultaneously. }

\textcolor{black}{However, we cannot simply quantize all variables
and parameters of a pre-trained neural network from floating-point
into fixed-point. Since quantization is a nonlinear operation, the
results would not be optimal for the fixed-point neural work. In addition,
direct quantization of variables and parameters may result in a loss
of gradients. Hereby, we propose a simulated quantization method to
train a neural network that can produce the optimal parameters in
fixed-point form. This is an essential step to prepare a CNN before
implement it on an FPGA.}

\textcolor{black}{}
\begin{figure}[h]
\begin{centering}
\textcolor{black}{}%
\begin{tabular}{cc}
\textcolor{black}{\includegraphics[width=0.4\columnwidth]{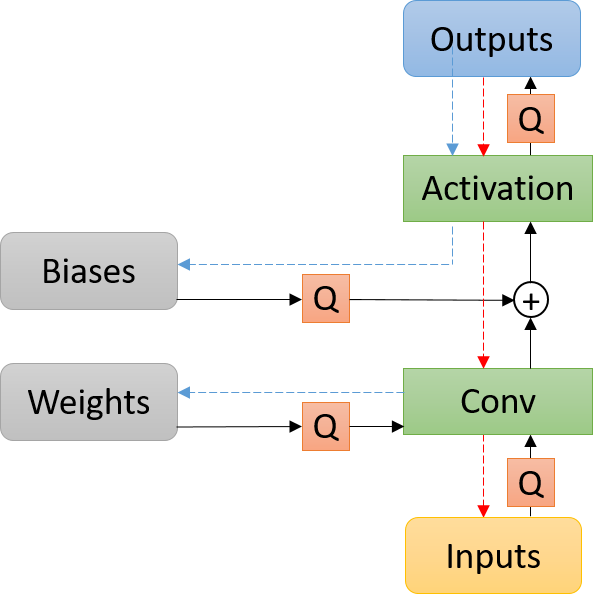}} &
\textcolor{black}{\includegraphics[width=0.4\columnwidth]{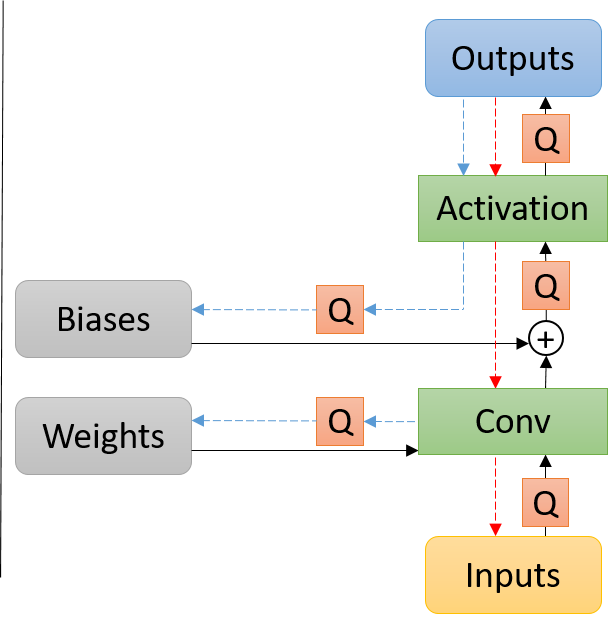}}\tabularnewline
\textcolor{black}{(a)} &
\textcolor{black}{(b)}\tabularnewline
\textcolor{black}{\includegraphics[width=0.4\columnwidth]{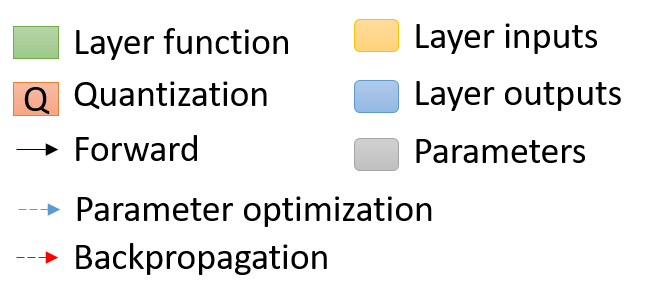}} &
\tabularnewline
\end{tabular}
\par\end{centering}
\textcolor{black}{\caption{(a) Simulated quantization method as in \cite{jacob_google2017quantization}
and (b) the proposed quantization method in this work\label{fig:quantization}}
}
\end{figure}

\textbf{\textcolor{black}{Simulated quantization of weights}}\textcolor{black}{:
Quantization that we refer here is not an simple operation of quantizing
all weights from floating-point to fixed-point numbers. Additional
training is needed to avoid negative impact on the accuracy. At the
training stage, however, floating-point weights are preferred because
we want to avoid gradient exploding and vanishing. In \cite{jacob_google2017quantization},
a simulated quantization approach was proposed in which weights and
gradients are stored as floating-point numbers during back-propagation
training but the quantized fixed-point numbers are used during forward
convolutional operations. The advantage was that the weights and gradients
are updated in continuous space so that local optimum due to quantization
can be avoided. The disadvantage was that several key functions need
to be modified to support this method. However, it is usually difficult
to modify, maintain and distribute customized components in a general
machine-learning platform such as TensorFlow. }

\textcolor{black}{In our work, a new weight regulator is defined and
added to the existing network. The regulator is described in }\textcolor{black}{\emph{Algorithm
}}\textcolor{black}{\ref{alg:qt_weight}. The key innovation is that
the regulator quantizes the weights during training and the fixed-point
numbers are used during forward operations. Meanwhile, the floating-point
weights are also stored in the memory that are used when computing
the gradients during back-propagation. However, quantization function
is not differentiable. Therefore, we introduce the }\textcolor{black}{\emph{StopGradient}}\textcolor{black}{{}
function. The }\textcolor{black}{\emph{StopGradient}}\textcolor{black}{{}
function is a built-in function in TensorFlow that force its gradient
to be zero for given input. By applying this function, the gradients
are kept the same as of floating point backpropagation, while the
weights are quantized. Hence, the proposed quantization algorithm
is imported to the TensorFlow platform as a plug-in regulator. The
proposed quantization algorithm and data flow are shown in Figure
\ref{fig:quantization} in comparison to the simulated quantization
in \cite{jacob_google2017quantization}.}

\textcolor{black}{}
\begin{algorithm}[h]
\textcolor{black}{\quad{}}\textbf{\textcolor{black}{Data Weights
W, Gradients G}}

\textcolor{black}{\quad{}}\textbf{\textcolor{black}{Parameter total\_bits=N,
fraction\_bits=F}}

\textcolor{black}{1: Fraction\_scalar $\leftarrow$ $2^{F}$}

\textcolor{black}{2: Upper\_bound $\leftarrow$$2^{N-1}-1$}

\textcolor{black}{3: Lower\_bound $\leftarrow$ $-2^{N-1}$}

\textcolor{black}{4: }\textbf{\textcolor{black}{$\hat{\boldsymbol{W}}$
$\leftarrow$$\boldsymbol{W}\times$}}\textcolor{black}{Fraction\_scalar}

\textcolor{black}{5: }\textbf{\textcolor{black}{$\hat{\boldsymbol{W}}$
$\leftarrow$}}\textcolor{black}{round(}\textbf{\textcolor{black}{$\hat{\boldsymbol{W}}$)}}

\textcolor{black}{6: }\textbf{\textcolor{black}{$\hat{\boldsymbol{W}}$
$\leftarrow$}}\textcolor{black}{max(}\textbf{\textcolor{black}{$\hat{\boldsymbol{W}}$}}\textcolor{black}{,Lower\_bound)}

\textcolor{black}{7: }\textbf{\textcolor{black}{$\hat{\boldsymbol{W}}$
$\leftarrow$}}\textcolor{black}{min(}\textbf{\textcolor{black}{$\hat{\boldsymbol{W}}$}}\textcolor{black}{,Upper\_bound)}

\textcolor{black}{8: }\textbf{\textcolor{black}{$\hat{\boldsymbol{W}}$
$\leftarrow$$\hat{\boldsymbol{W}}$}}\textcolor{black}{/Fraction\_scalar}

\textcolor{black}{9:}\textbf{\textcolor{black}{{} $\boldsymbol{W}$
$\leftarrow$$\boldsymbol{W}$ + StopGradient($\hat{\boldsymbol{W}}-\boldsymbol{W}$)}}

\textcolor{black}{10: }\textbf{\textcolor{black}{return}}\textcolor{black}{{}
\{}\textbf{\textcolor{black}{$\boldsymbol{W}\boldsymbol{,G}$\}}}
\begin{description}
\item [{\caption{weight quantization\label{alg:qt_weight}}
}]~
\end{description}
\end{algorithm}

\textbf{\textcolor{black}{Simulated quantization of variables:}}\textcolor{black}{{}
Quantization of variables is similar to the weights. Quantization
of variables is implemented as a new activation function so that it
can be imported as a custom defined function rather than modifying
the body of an existing platform. As described in}\textcolor{black}{\emph{
Algorithm }}\textcolor{black}{\ref{alg:qt_weight}, if we convert
a floating-point number to a }\textcolor{black}{\emph{N}}\textcolor{black}{-bit
fixed-point number with an }\textcolor{black}{\emph{F}}\textcolor{black}{-bit
fraction, the operation is to shift to the left by }\textcolor{black}{\emph{F}}\textcolor{black}{{}
bits and then round it to a }\textcolor{black}{\emph{N}}\textcolor{black}{-bit
integer, followed by shifting back $F$ bits to the right. To minimize
negative impact on the back-propagation training, the gradients are
all computed in floating-point.}

\textbf{\textcolor{black}{Evaluation of quantization: }}\textcolor{black}{To
evaluate the influence of quantization on accuracy, we first trained
the ChipNet in floating point using the Ford training set, and then
quantized and fine-tuned using the same training set. Both versions
of the ChipNet with and without quantization are evaluated on the
same dataset. The result listed in Table \ref{tab:Result_Ford} shows
that ChipNet quantized to 18 or more bits has similar performance
compared to the floating-point model, which indicates that our proposed
quantization scheme does not cause accuracy degradation for convolutional
neural networks.}

\subsection{\textcolor{black}{View of drivable region}}

\textcolor{black}{The output of network denotes the possibility of
drivable region for each cell in spherical view. In post-processing,
the output is projected to the top view of a 20-meter wide and 40-meter
long area in front of the vehicle. We choose topview in post-processing
because it matches the output data format in KITTI benchmark \cite{Fritsch_KIT2013KITTI_road},
so we can compare our results with others reported in the dataset. }

\textcolor{black}{The post-processing algorithm is described in Algorithm
\ref{alg:post-processing}. Suppose the possibility threshold of a
drivable region is set to $THR$, then the reference point in each
column $j$ in the network output ${P_{i,j}}$ is determined by the
nearest LiDAR point in group $\left\{ P|_{col=j,\:p<THR}\right\} $.
After generating the reference points, a contour of the drivable region
becomes a polygon that contains all reference points as vertices.}

\textcolor{black}{The post-processing scheme is implemented on CPU
using GridMap\cite{Fankhauser2016GridMapLibrary} that is an universal
grid map management library. The GridMap library stores map data as
Eigen matrix and supports iterators for rectangular, circular, polygonal
regions and lines allowing convenient and efficient cell data access.
In post-processing, we initialize a grid map instance with a range
setting of {[}6, 46{]} meters in }\textcolor{black}{\emph{x}}\textcolor{black}{-coordinate
and {[}-10, 10{]} meters in }\textcolor{black}{\emph{y}}\textcolor{black}{-coordinate.
The resolution is set to 0.05 meter per cell so that the grid map
has 800 cells in x-coordinate and 400 cells in y-coordinate. When
the post-processing node receives a network output frame, it stores
the frame as an Eigen matrix. In }\textcolor{black}{\emph{Algorithm
\ref{alg:post-processing}}}\textcolor{black}{, Step 1-3 is processed
on the matrix. In Step 4-5, the contour vertices are imported to a
polygon iterator instance and then the drivable region is labeled
cell by cell as the polygon iterates. The execution time of post-processing
is 5ms per frame on a typical CPU.}

\textcolor{black}{The post-processing affects the segmentation accuracy
in two parts: (1) After projected to the top view, LiDAR scans are
so sparse in far range that there exists distortion between the projected
LiDAR boundary and the real drivable region boundary, and (2) Algorithm
\ref{alg:post-processing} assumes that the space inside the polygon
is drivable while the space outside is not, so there exists an error
if the drivable region has holes. We use the KITTI training set to
evaluate the effect quantitatively. By projecting the ground truth
in Spherical view to top view using the post-processing algorithm
and comparing with the top view ground  truth, we found that the F1
- score is limited to 95.5\% by the post-processing algorithm.}

\textcolor{black}{For visualization purpose, we also generate a drivable
region map on camera view by applying a similar post-processing procedure.}

\textcolor{black}{}
\begin{algorithm}[h]
\textcolor{black}{$\quad$}\textbf{\textcolor{black}{Data Input tensor
I}}\textcolor{black}{=$\{I_{i,j}\}$}\textbf{\textcolor{black}{, output
tensor P}}\textcolor{black}{=$\{P_{i,j}\}$}

\textcolor{black}{$\quad$}\textbf{\textcolor{black}{Parameter threshold
}}\textcolor{black}{$THR$.}

\textcolor{black}{1: $\boldsymbol{I}\leftarrow Threshold(\boldsymbol{I}\geqslant THR)$}

\textcolor{black}{2: }\textbf{\textcolor{black}{$\boldsymbol{\hat{I}}\leftarrow GetLargestConnectedComponent(\boldsymbol{I})$}}

\textcolor{black}{3: $\boldsymbol{\hat{I}}\leftarrow Dilation(\boldsymbol{\hat{I}},disk(1)$)}

\textcolor{black}{4: $\boldsymbol{B}\leftarrow GetContour(\boldsymbol{\hat{I}})$}

\textcolor{black}{5: $\boldsymbol{\widetilde{B}}\leftarrow ProjectToTargetView(\boldsymbol{B})$}

\textcolor{black}{5: $\boldsymbol{\widetilde{A}}\leftarrow Polygon(\boldsymbol{\widetilde{B}})$}

\textcolor{black}{6: }\textbf{\textcolor{black}{return $\boldsymbol{\widetilde{A}}$}}

\textcolor{black}{\caption{Post-processing of CNN output as segmentation results\label{alg:post-processing}}
}
\end{algorithm}

\subsection{\textcolor{black}{Network training and evaluation}}

\textcolor{black}{The training platform of ChipNet is a workstation
with Xeon 2.4 GHz CPU and NVidia K20 GPU. The software environment
is a Python based framework named Keras \cite{chollet2015keras} with
TensorFlow 1.4 back end. The input of the network is an $180\times64\times14$
tensor and the output of the network is an $180\times64\times1$ tensor.
The training speed on the platform is $256$ ms per frame. To evaluate
the performance of the proposed solution, a subset of the Ford Campus
Vision and LiDAR Dataset\cite{pandey2011ford} and the KITTI road
benchmark\cite{Fritsch_KIT2013KITTI_road} is used for training and
testing purposes.}

\textcolor{black}{The original Ford Dataset described in \cite{pandey2011ford}
contains 3871 frames of LiDAR data recorded with synchronized camera
data. The LiDAR data are sampled at 10 Hz. The dataset itself has
no labels or annotations, so we created a subset and labeled the drivable
region manually. To reduce the overlaps among the consecutive frames,
we selected only 1 out of every 5 consecutive frames. Effectively
the dataset is downsampled to 2 frames per second. We also removed
some off-road samples, such as vehicles on the parking lot, from the
dataset since we concentrated on road scenarios. Therefore, we generated
a 600-frame subset from the Ford dataset for training and evaluation. }

\textcolor{black}{In the subset, the original image is cropped from
the size of $1243\times1616$ to $800\times200$ resolution that overlaps
with the LiDAR point cloud. The data is arranged as described in Section
\ref{subsec:LiDAR-data-preparation}. In order to obtain the LiDAR
ground truth, a ray tracing approach is applied as described in }\textcolor{black}{\emph{Algorithm}}\textcolor{black}{{}
\ref{alg:LiDAR_gt}. The projection method from LiDAR coordinate to
camera coordinate is described in }\textcolor{black}{\emph{Algorithm}}\textcolor{black}{{}
\ref{alg:LiDAR_to_camera}. In our labeled subset, each sample includes
a $180\times64\times14$ LiDAR frame, a $800\times200$ color image,
a $180\times64\times1$ LiDAR ground truth frame and a $800\times200$
ground truth image. We randomly selected 400 samples for training/validation
and the remaining 200 samples for evaluation. Furthermore, we augmented
the training samples through rotating the field of view by $(-10,-5,0,5,10)$
degrees from the LiDAR ground truth. Thus, we generated a training
set with $2000$ samples. }

\textcolor{black}{Cross entropy was selected as the loss function
and Adam \cite{kingma2014adam} method with default settings was selected
as the optimizer. We first trained the network without the quantization
plug-in for 30 epochs, at which time the training process converged
well. We then fine-tuned the network with the quantization plug-in
for 10 epochs to obtain the fixed-point weights. The initial training
took 4.5 hours and the fine-turning took 1.5 hours. For each defined
fixed-point bit-length format, we applied the same simulated quantization
procedure during fine-tuning. The bit length resulted the least loss
is chosen for the FPGA implementation.}

\textcolor{black}{}
\begin{algorithm}[h]
\textcolor{black}{$\quad$}\textbf{\textcolor{black}{Data Input tensor
L}}\textcolor{black}{=$\{\boldsymbol{L_{i,j,k}}\}$}\textbf{\textcolor{black}{,
Ground truth image B=}}\textcolor{black}{$\{\boldsymbol{B_{i,j}}\}$}

\textcolor{black}{1: }\textbf{\textcolor{black}{$\boldsymbol{\boldsymbol{\overrightarrow{x_{1}}}}\leftarrow\left[\boldsymbol{\begin{array}{c}
L_{i,j,1}\\
L_{i,j,2}\\
L_{i,j,3}
\end{array}}\right]$, $\boldsymbol{\boldsymbol{\overrightarrow{x_{2}}}}\leftarrow\left[\begin{array}{c}
\boldsymbol{L_{i,j,8}}\\
\boldsymbol{L_{i,j,9}}\\
\boldsymbol{L_{i,j,10}}
\end{array}\right]$}}

\textcolor{black}{2: $\boldsymbol{\boldsymbol{\hat{\overrightarrow{x_{1}}}}}\leftarrow Proj(\boldsymbol{\boldsymbol{\overrightarrow{x_{1}}}})$,
$\boldsymbol{\boldsymbol{\hat{\overrightarrow{x_{2}}}}}\leftarrow Proj(\boldsymbol{\boldsymbol{\overrightarrow{x_{2}}}})$}

\textcolor{black}{3: $\boldsymbol{G_{i,j,1}}\leftarrow[\boldsymbol{B}(\boldsymbol{\hat{\overrightarrow{x_{1}}}})>0]\times[\boldsymbol{B}(\boldsymbol{\hat{\overrightarrow{x_{2}}}})>0]$}

\textcolor{black}{4: }\textbf{\textcolor{black}{return }}\textcolor{black}{$\boldsymbol{G}$}

\textcolor{black}{\caption{Ground truth labeling for LiDAR samples\label{alg:LiDAR_gt}}
}
\end{algorithm}

\textcolor{black}{}
\begin{algorithm}[h]
\textcolor{black}{$\quad$}\textbf{\textcolor{black}{Data LiDAR point
$P{x_{i},y_{i},z_{i},r_{i}}$}}

\textcolor{black}{$\quad$}\textbf{\textcolor{black}{Parameter transform
matrix }}\textcolor{black}{$\boldsymbol{K}$$\in R^{3\times4}$}

\textcolor{black}{1: }\textbf{\textcolor{black}{$\left[\begin{array}{c}
\hat{x_{i}}\\
\hat{y_{i}}\\
\hat{z_{i}}
\end{array}\right]\leftarrow\boldsymbol{K}\left[\begin{array}{c}
x_{i}\\
y_{i}\\
z_{i}\\
1
\end{array}\right]$}}

\textcolor{black}{2: $\left[\begin{array}{c}
\hat{x_{i}}\\
\hat{y_{i}}
\end{array}\right]\leftarrow\left[\begin{array}{c}
\frac{\hat{x_{i}}}{\hat{z_{i}}}\\
\frac{\hat{y_{i}}}{\hat{z_{i}}}
\end{array}\right]$}

\textcolor{black}{3: }\textbf{\textcolor{black}{return $\left[\begin{array}{c}
\hat{x_{i}}\\
\hat{y_{i}}
\end{array}\right]$}}

\textcolor{black}{\caption{Projection from LiDAR coordinate to camera coordinate\label{alg:LiDAR_to_camera}}
}
\end{algorithm}

\textcolor{black}{In the testing session, we selected F1 score (F1),
average precision (AP), precision (PRE), recall (REC), false positive
rate (FPR) and false negative rate (FNR) in image view as the evaluating
metrics. The metrics are computed as in (1-4). Table \ref{tab:Result_Ford}
presents the evaluation results using different bit length of fixed-point
quantization. The result shows that the proposed network quantized
to 16 or more bits has comparable accuracy to floating-point results,
but accuracy drops sharply if quantization is below 16 bits. In our
work, 18 bits are selected since it is the best choice supported by
the target FPGA platform.}

\textcolor{black}{
\begin{equation}
Precision=\frac{TP}{TP+FP}
\end{equation}
}

\textcolor{black}{
\begin{equation}
Recall=\frac{TP}{TP+FN}
\end{equation}
}

\textcolor{black}{
\begin{equation}
F1\;score=\frac{2\cdot Precision\cdot Recall}{Precision+Recall}
\end{equation}
}

\textcolor{black}{
\begin{equation}
AP=\frac{TP+TN}{TP+FP+TN+FN}
\end{equation}
}

\textcolor{black}{We also evaluate our network in KITTI road benchmark
\cite{Fritsch_KIT2013KITTI_road}. The KITTI vision benchmark suite
is a widely used dataset that contains LiDAR, camera, GPS and IMU
data. In addition, the vertex transformation from LiDAR coordinate
to camera coordinate is provided. The road benchmark in the suite
includes $289$ training samples and $290$ test samples. The point
cloud was acquired by a 64-line Velodyne laser scanner and the camera
frames were recorded from a Point Grey 1.4 megapixels camera. For
better sensor fusion, the LiDAR point cloud is rectified at each time
step and the camera frame is cropped to a $375\times1242$ image.
In addition, the data frames from different sensors are synchronized
to 10 Hz. }

\textcolor{black}{Different from the Ford dataset that evaluates on
camera view, the KITTI road benchmark evaluates the segmentation results
on top view, in which the result is mapped to a $400\times800$ image.
The mapped image represents the accessibility of the region of 40
meters in the front (from 6 meters to 46 meters) and 10 meters on
each side (left and right).}

\textcolor{black}{In the training session, we first augmented the
dataset through rotating the field of view by $(\lyxmathsym{\textendash}10,-8,-6,-4,-2,0,2,4,6,8,10)$
degrees from the corresponding LiDAR ground truth. So, we obtained
$3179$ samples, among that $3000$ randomly selected samples are
used for training and the other $179$ samples are used for validation.
We fine-tuned the network with quantization plug-in for 10 epochs
from the weights trained in Ford dataset, and submitted the results
to the benchmark online evaluator. The training time was $2.05$ hours. }

\begin{figure}
\begin{centering}
\textcolor{black}{}%
\begin{tabular}{cc}
\textcolor{black}{\includegraphics[width=0.47\columnwidth]{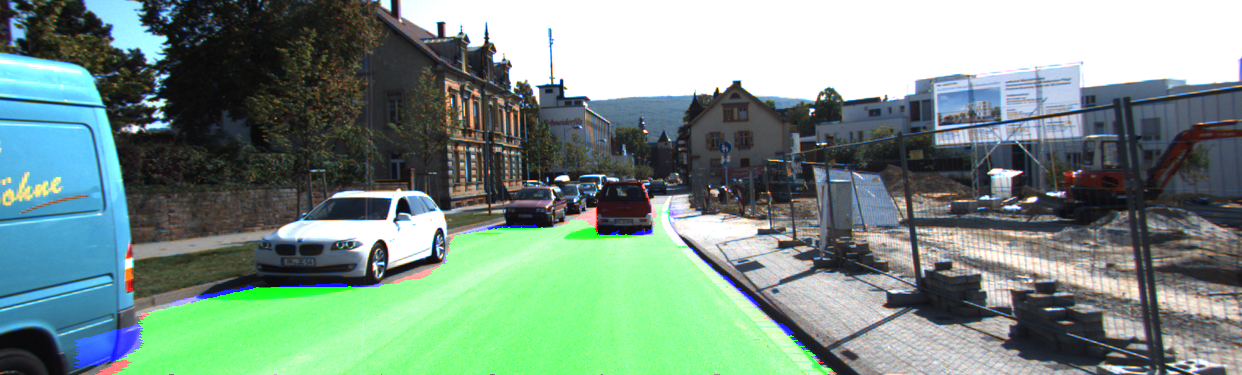}} &
\textcolor{black}{\includegraphics[width=0.47\columnwidth]{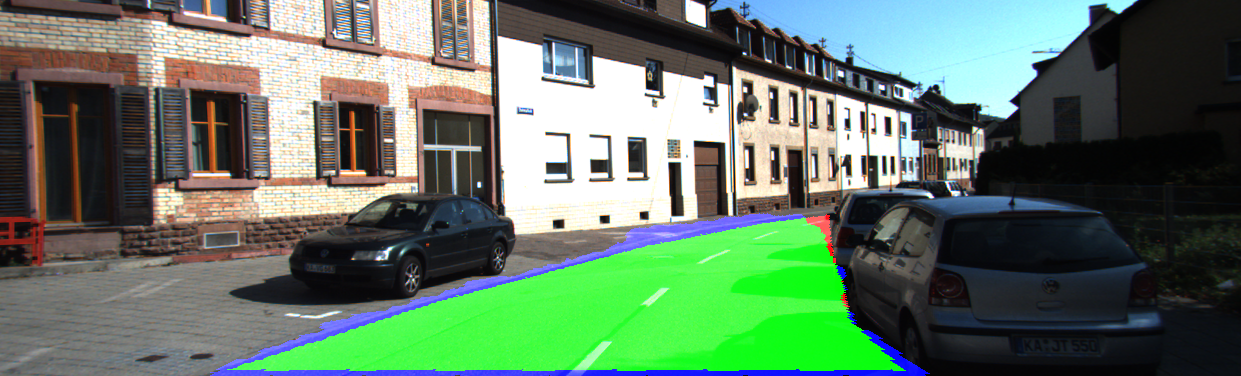}}\tabularnewline
\textcolor{black}{(a)} &
\textcolor{black}{(b)}\tabularnewline
\textcolor{black}{\includegraphics[width=0.47\columnwidth]{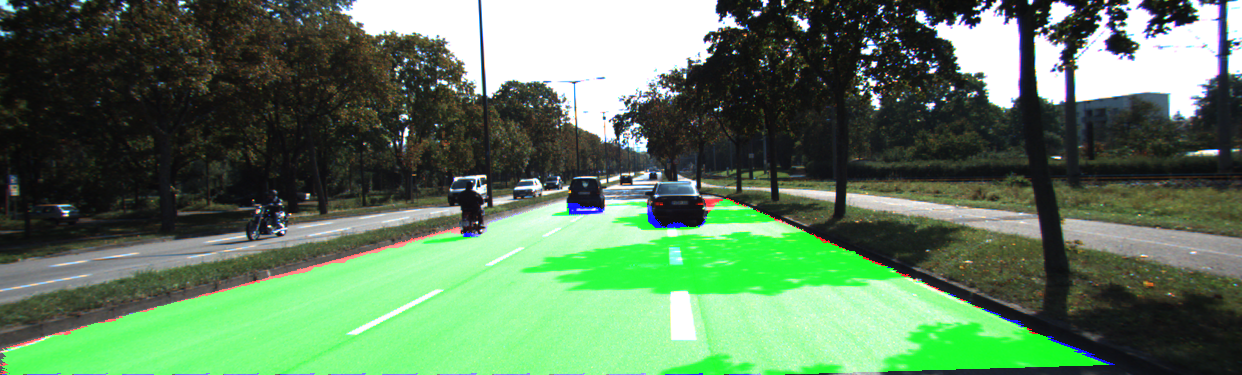}} &
\textcolor{black}{\includegraphics[width=0.47\columnwidth]{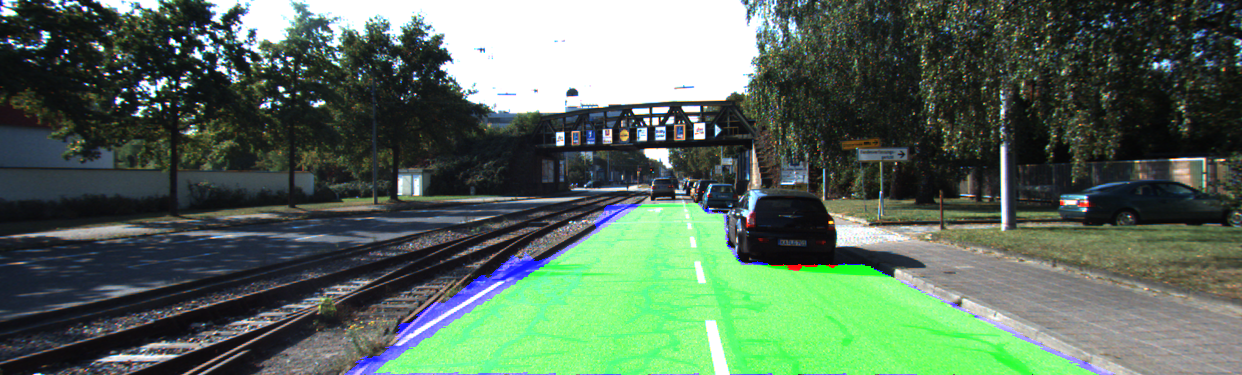}}\tabularnewline
\textcolor{black}{(c)} &
\textcolor{black}{(d)}\tabularnewline
\end{tabular}
\par\end{centering}
\textcolor{black}{\caption{Examples of segmentation results from KITTI road dataset \label{fig:Results_KITTI}}
}
\end{figure}

\textcolor{black}{A comparison with several existing results is presented
in Table \ref{tab:Result_KITTI}. Typical results are shown in Figure
\ref{fig:Results_KITTI}. The red area denotes the false drivable
region (false positive), the blue area denotes the missing driving
region (false negative), the green area denotes the correct drivable
region (true positive), and the rest area denotes the correct forbidden
region (true negative) or don\textquoteright t care region. We also
evaluated ChipNet on the front view as in \cite{chen_baidu2017multi_3d}
without quantization, which requires significant more run time but
results lower accuracy than the spherical view. It implies that LiDAR
data arrangement spherical view reveals more features as input to
the CNN.}

\textcolor{black}{Our proposed approach can provide highly reliable
drivable region segmentation with minor distortions around the road
boundary. For vehicles on the road, the segmentation boundary matches
the ground truth boundary or slightly distorts towards the road center
that is safe for automated driving. For the road with sidewalk, the
segment boundary matches the ground truth if the sidewalk is above
the road surface. However, if the sidewalk is equal or below the road
surface, the detected drivable region sometimes extends 1 to 2 meters
into the sidewalk, which needs to be improved in future research.
In addition, our solution returns accurate drivable regions in poor
illumination scenarios such as inside tunnels or facing the sun glare.
In contract, vision based solutions rarely work well in those scenarios. }

\textcolor{black}{}
\begin{table}[h]
\begin{centering}
\textcolor{black}{\caption{Performance impact of quantization evaluated on Ford dataset\label{tab:Result_Ford}}
}
\par\end{centering}
\centering{}\textcolor{black}{}%
\begin{tabular}{|>{\centering}m{0.2\columnwidth}|>{\centering}m{0.07\columnwidth}|>{\centering}m{0.07\columnwidth}|>{\centering}m{0.07\columnwidth}|>{\centering}m{0.07\columnwidth}|>{\centering}m{0.07\columnwidth}|>{\centering}m{0.07\columnwidth}|}
\hline 
\textcolor{black}{Name} &
\textcolor{black}{F1 $\:$\%} &
\textcolor{black}{AP \%} &
\textcolor{black}{PRE \%} &
\textcolor{black}{REC \%} &
\textcolor{black}{FPR \%} &
\textcolor{black}{FNR \%}\tabularnewline
\hline 
\textcolor{black}{ChipNet}

\textcolor{black}{without }

\textcolor{black}{Quantization} &
\textcolor{black}{86.6} &
\textcolor{black}{94.0} &
\textcolor{black}{85.5} &
\textcolor{black}{87.6} &
\textcolor{black}{14.5} &
\textcolor{black}{3.5}\tabularnewline
\hline 
\textcolor{black}{ChipNet}

\textcolor{black}{Quantized to 32 bit} &
\textcolor{black}{86.7} &
\textcolor{black}{94.2} &
\textcolor{black}{86.7} &
\textcolor{black}{86.7} &
\textcolor{black}{13.3} &
\textcolor{black}{3.7}\tabularnewline
\hline 
\textcolor{black}{ChipNet}

\textcolor{black}{Quantized to 24 bit} &
\textcolor{black}{86.9} &
\textcolor{black}{94.1} &
\textcolor{black}{85.1} &
\textcolor{black}{88.8} &
\textcolor{black}{14.9} &
\textcolor{black}{3.2}\tabularnewline
\hline 
\textcolor{black}{ChipNet}

\textcolor{black}{Quantized to 18 bit} &
\textcolor{black}{86.3} &
\textcolor{black}{94.0} &
\textcolor{black}{86.0} &
\textcolor{black}{86.7} &
\textcolor{black}{14.0} &
\textcolor{black}{3.7}\tabularnewline
\hline 
\textcolor{black}{ChipNet}

\textcolor{black}{Quantized to 16 bit} &
\textcolor{black}{86.9} &
\textcolor{black}{93.6} &
\textcolor{black}{81.8} &
\textcolor{black}{90.8} &
\textcolor{black}{18.2} &
\textcolor{black}{2.6}\tabularnewline
\hline 
\textcolor{black}{ChipNet}

\textcolor{black}{Quantized to 12 bit} &
\textcolor{black}{83.7} &
\textcolor{black}{92.3} &
\textcolor{black}{78.0} &
\textcolor{black}{90.3} &
\textcolor{black}{22.0} &
\textcolor{black}{2.8}\tabularnewline
\hline 
\textcolor{black}{SegNet\cite{badrinarayanan2017segnet}} &
\textcolor{black}{89.2} &
\textcolor{black}{93.6} &
\textcolor{black}{89.4} &
\textcolor{black}{88.9} &
\textcolor{black}{10.5} &
\textcolor{black}{4.6}\tabularnewline
\hline 
\end{tabular}
\end{table}

\textcolor{black}{}
\begin{figure}
\begin{centering}
\textcolor{black}{}%
\begin{tabular}{cc}
\textcolor{black}{\includegraphics[width=0.47\columnwidth]{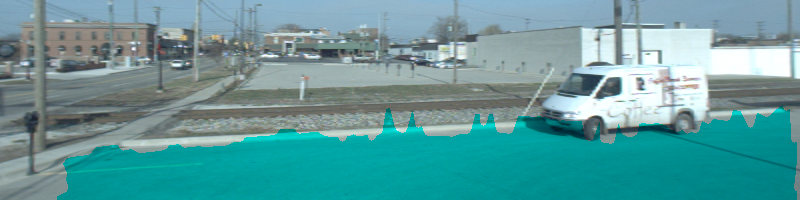}} &
\textcolor{black}{\includegraphics[width=0.47\columnwidth]{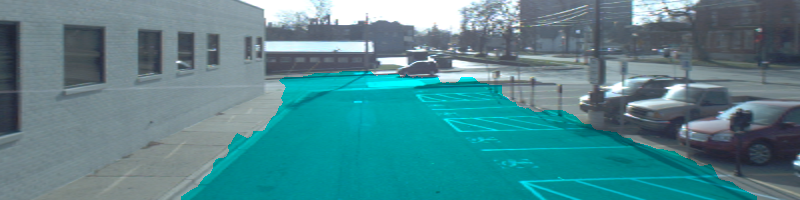}}\tabularnewline
\textcolor{black}{(a)} &
\textcolor{black}{(b)}\tabularnewline
\textcolor{black}{\includegraphics[width=0.47\columnwidth]{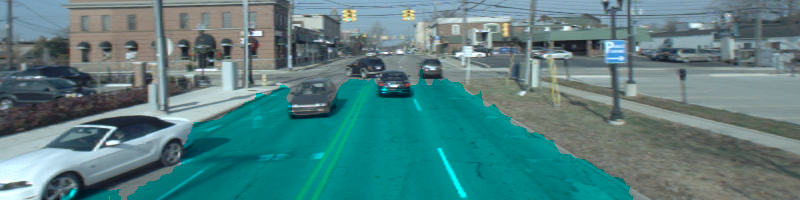}} &
\textcolor{black}{\includegraphics[width=0.47\columnwidth]{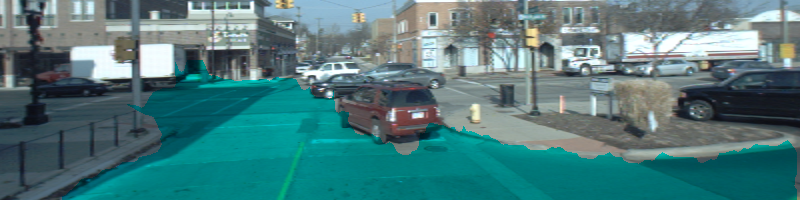}}\tabularnewline
\textcolor{black}{(c)} &
\textcolor{black}{(d)}\tabularnewline
\end{tabular}
\par\end{centering}
\textcolor{black}{\caption{Examples of the segmentation results from Ford dataset\label{fig:Results_Ford} }
}
\end{figure}

\textcolor{black}{}
\begin{table*}
\begin{centering}
\textcolor{black}{\caption{Comparison with existing results on KITTI road benchmark\label{tab:Result_KITTI}}
}
\par\end{centering}
\centering{}\textcolor{black}{}%
\begin{tabular}{|c|c|c|c|c|c|c|c|}
\hline 
\textcolor{black}{Name} &
\textcolor{black}{F1 \%} &
\textcolor{black}{AP \%} &
\textcolor{black}{PRE \%} &
\textcolor{black}{REC \%} &
\textcolor{black}{FPR \%} &
\textcolor{black}{FNR \%} &
\textcolor{black}{Runtime ms}\tabularnewline
\hline 
\textcolor{black}{ChipNet on FPGA (this work)} &
\textcolor{black}{94.05} &
\textcolor{black}{88.29} &
\textcolor{black}{93.57} &
\textcolor{black}{94.53} &
\textcolor{black}{3.58} &
\textcolor{black}{5.47} &
\textcolor{black}{17.59}\tabularnewline
\hline 
\textcolor{black}{LoDNN\cite{caltagirone2017LoDNN}} &
\textcolor{black}{94.07} &
\textcolor{black}{92.03} &
\textcolor{black}{92.81} &
\textcolor{black}{95.37} &
\textcolor{black}{4.07} &
\textcolor{black}{4.63} &
\textcolor{black}{18}\tabularnewline
\hline 
\textcolor{black}{HybridCRF\cite{Xiao2017HybridCRF}} &
\textcolor{black}{90.81} &
\textcolor{black}{86.01} &
\textcolor{black}{91.05} &
\textcolor{black}{90.57} &
\textcolor{black}{4.90} &
\textcolor{black}{9.43} &
\textcolor{black}{1500}\tabularnewline
\hline 
\textcolor{black}{LidarHisto\cite{chen2017lidar_histogram+RANSAC}} &
\textcolor{black}{90.67} &
\textcolor{black}{84.79} &
\textcolor{black}{93.06} &
\textcolor{black}{88.41} &
\textcolor{black}{3.63} &
\textcolor{black}{11.59} &
\textcolor{black}{100}\tabularnewline
\hline 
\textcolor{black}{MixedCRF\cite{Han2017MixedCRF}} &
\textcolor{black}{90.59} &
\textcolor{black}{84.24} &
\textcolor{black}{89.11} &
\textcolor{black}{92.13} &
\textcolor{black}{6.20} &
\textcolor{black}{7.87} &
\textcolor{black}{6000}\tabularnewline
\hline 
\textcolor{black}{FusedCRF\cite{Xiao2015FusedCRF}} &
\textcolor{black}{88.25} &
\textcolor{black}{79.24} &
\textcolor{black}{83.62} &
\textcolor{black}{93.44} &
\textcolor{black}{10.08} &
\textcolor{black}{6.56} &
\textcolor{black}{2000}\tabularnewline
\hline 
\textcolor{black}{RES3D-Velo\cite{Shinzato2014RES3D-Velo}} &
\textcolor{black}{86.58} &
\textcolor{black}{78.34} &
\textcolor{black}{82.63} &
\textcolor{black}{90.92} &
\textcolor{black}{10.53} &
\textcolor{black}{9.08} &
\textcolor{black}{60}\tabularnewline
\hline 
\textcolor{black}{ChipNet on camera view} &
\textcolor{black}{82.50} &
\textcolor{black}{86.13} &
\textcolor{black}{77.37} &
\textcolor{black}{88.36} &
\textcolor{black}{14.23} &
\textcolor{black}{11.64} &
\textcolor{black}{945}\tabularnewline
\hline 
\end{tabular}
\end{table*}

\textcolor{black}{}

\section{\textcolor{black}{Hardware Architecture\label{sec:Hardware-Architecture}}}

\begin{figure}
\begin{centering}
\textcolor{black}{\includegraphics[height=4cm]{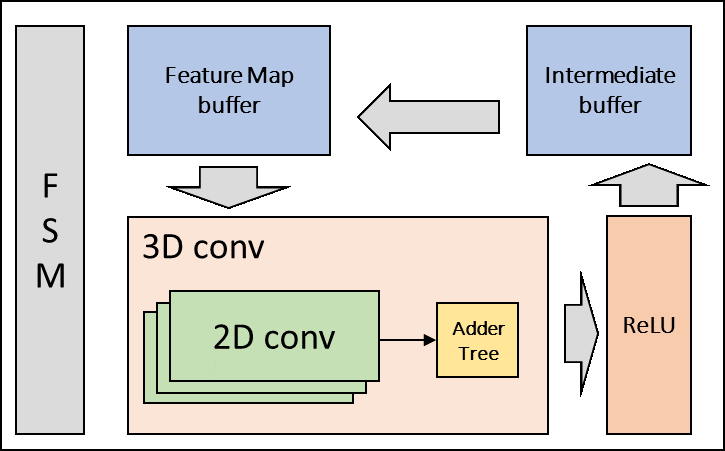}}
\par\end{centering}
\textcolor{black}{\caption{Hardware architecture of ChipNet convolutional neural network\label{fig:Hardware-architecture}}
}
\end{figure}

\textcolor{black}{As described in Section III, the LiDAR data after
pre-processing has 14 channels and the input data size is $180\times64$.
After the first layer of convolutional encoding, it becomes a feature
map with 64 channels. In the next 10 convolution layers, the input
and output feature map sizes remain the same as $180\times64\times64$.
The final layer performs the channel-wise mapping that produces an
output map of $180\times64$, each indicating the possibility of drivable
regions. The block diagram of hardware architecture is illustrated
in Figure \ref{fig:Hardware-architecture}. The system consists of
a 3D convolution unit, a ReLU block, a feature map buffer and an intermediate
buffer. 2D convolution and adder trees are embedded in the 3D convolution
block. Since the feature maps in each stage of ChipNet have the same
size, this 3D convolution unit is used repetitively. A finite state
machine (FSM) is designed to control the iterative processing steps.}

\textcolor{black}{}

\subsection{\textcolor{black}{Zero padding}}

\begin{figure}
\begin{centering}
\textcolor{black}{\includegraphics[height=4.5cm]{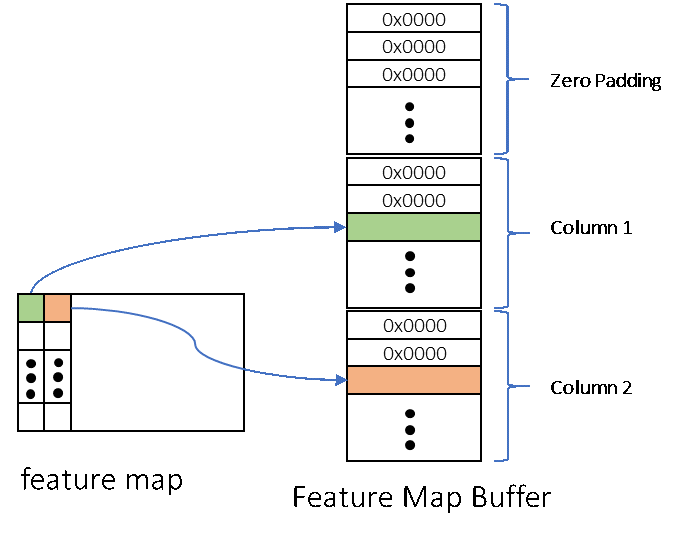}}
\par\end{centering}
\textcolor{black}{\caption{Automatic implementation of zero-padding in hardware\label{fig:zero-padding} }
}
\end{figure}

\textcolor{black}{In order to properly process the information along
the boundaries, zero padding must be applied to the feature map produced
by the convolution layer output. In our system, a dual-port RAM is
implemented for automatic zero-padding. In Figure \ref{fig:zero-padding},
all memory locations are pre-loaded with zeroes. Pixels of a feature
map are written to the corresponding address locations in the feature
map buffer. When reading the feature map from the feature map buffer
in continuous addresses, data are automatically zero-padded. The RAM
functions as the feature map buffer.}

\textcolor{black}{}

\subsection{\textcolor{black}{Convolution}}

\begin{figure}
\begin{centering}
\textcolor{black}{\includegraphics[height=4cm]{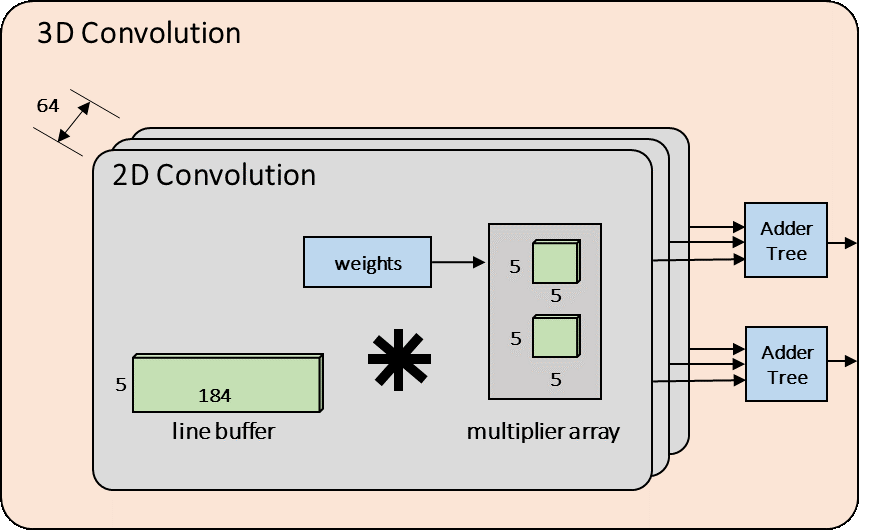}}
\par\end{centering}
\textcolor{black}{\caption{Block diagram of 3D and 2D convolution unit\label{fig:Hardware-convolution}}
}
\end{figure}

\begin{figure}
\begin{centering}
\textcolor{black}{\includegraphics[scale=0.4]{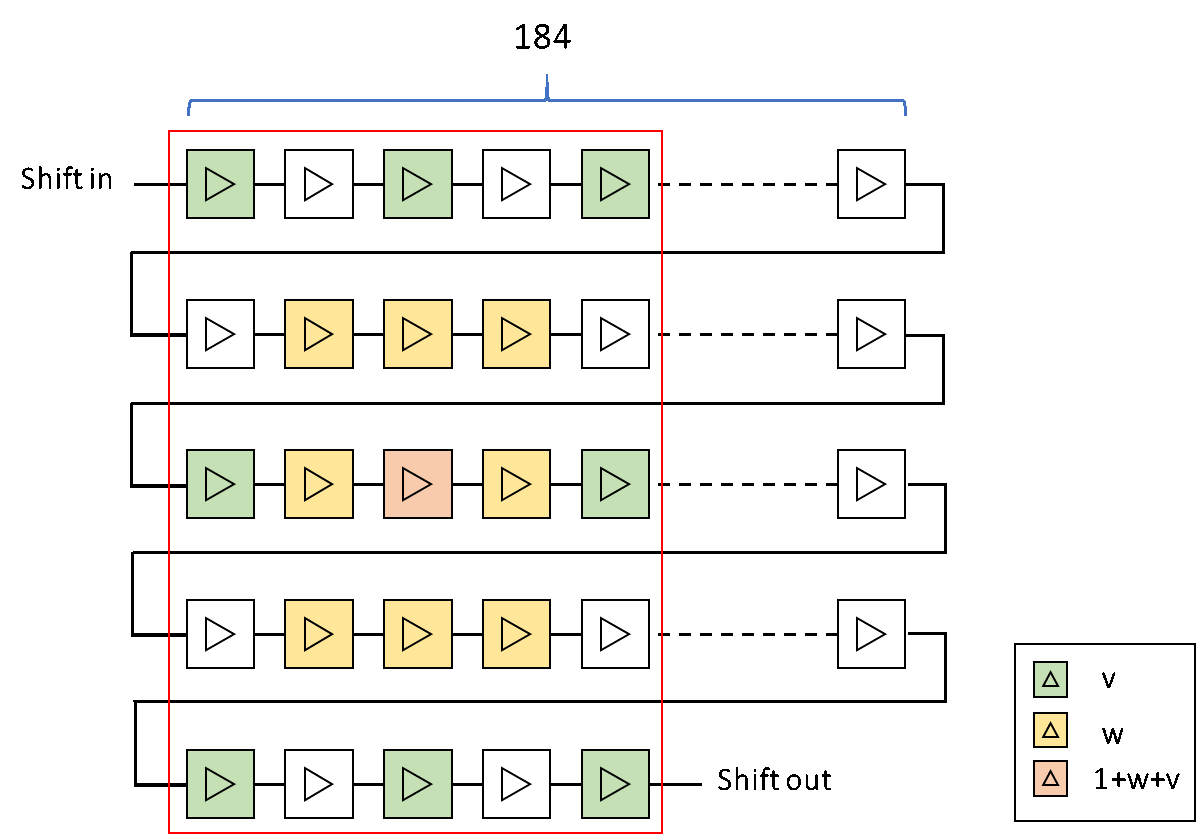}}
\par\end{centering}
\textcolor{black}{\caption{Block diagram of the line buffer unit\label{fig:Line-buffer}}
}
\end{figure}

\textcolor{black}{As exhibited in Figure \ref{fig:Hardware-convolution},
the 3D convolution unit contains 64 pieces of 2D convolution slices.
Each 2D convolution slice is built with a line buffer and two $5\times5$
multiplier arrays. The line buffer is designed using shift registers
as shown in Figure \ref{fig:Line-buffer}. It outputs a $5\times5$
window (outlined in red) as the input to the multiplier arrays. The
registers in green multiply with the dilated $3\times3$ convolution
kernel, and registers in yellow multiply with the regular $3\times3$
convolution kernel, and the register at the center multiplies with
the coefficient sum as shown in Figure \ref{fig:Block}. }

\textcolor{black}{In each 2D convolution block, the input data are
fed from the line buffer to two multiplier arrays, each followed by
an adder tree. The 2D convolution block is a pipeline architecture
that can process two convolution kernel operations in parallel. Since
each 2D convolution operation has 64 convolution kernels, the same
feature map is reloaded and processed for 32 times. All weights are
stored in on-chip memory to avoid the latency of off-chip memory access.
The ReLU block is implemented by a comparator and a multiplexer. If
the input value is larger than 0, it outputs the original value. Otherwise
the ReLU block outputs 0.}

\textcolor{black}{}

\textcolor{black}{}

\subsection{\textcolor{black}{FSM Controller}}

\textcolor{black}{Since the multiplier array in Figure \ref{fig:Hardware-convolution}
consumes a large number of DSP slices on FPGA, reusing it for each
convolution layer is a key consideration in the control logic design.
Thus, a cascaded finite state machine (FSM) is deployed to control
the iterative process. As shown in Figure \ref{fig:Hardware-convolution},
the 3D convolution unit can perform 2 kernel operations in parallel.
From Table \ref{tab:ChipNet-architecture}, each layer requires 64
convolutional kernel operations. So, the inner FSM controls the 3D
convolution unit to perform the same operations 32 times, each with
different input of feature maps, while the outer FSM controls the
order of layers.}

\textcolor{black}{As is shown in Figure \ref{fig:Hardware-architecture},
during each layer of convolution, the outer FSM first loads the input
feature map into feature map buffer. Meanwhile, the inner FSM starts
to feed feature map into the 3D convolution unit. The intermediate
feature maps are stored in intermediate buffer. When the inner FSM
completes the convolution of one layer, the outer FSM moves the data
from intermediate buffer to feature map buffer, and then it starts
the convolution of the next layer.}

\textcolor{black}{}%

\subsection{\textcolor{black}{Implementation Results}}

\begin{figure}
\textcolor{black}{\includegraphics[width=0.8\columnwidth]{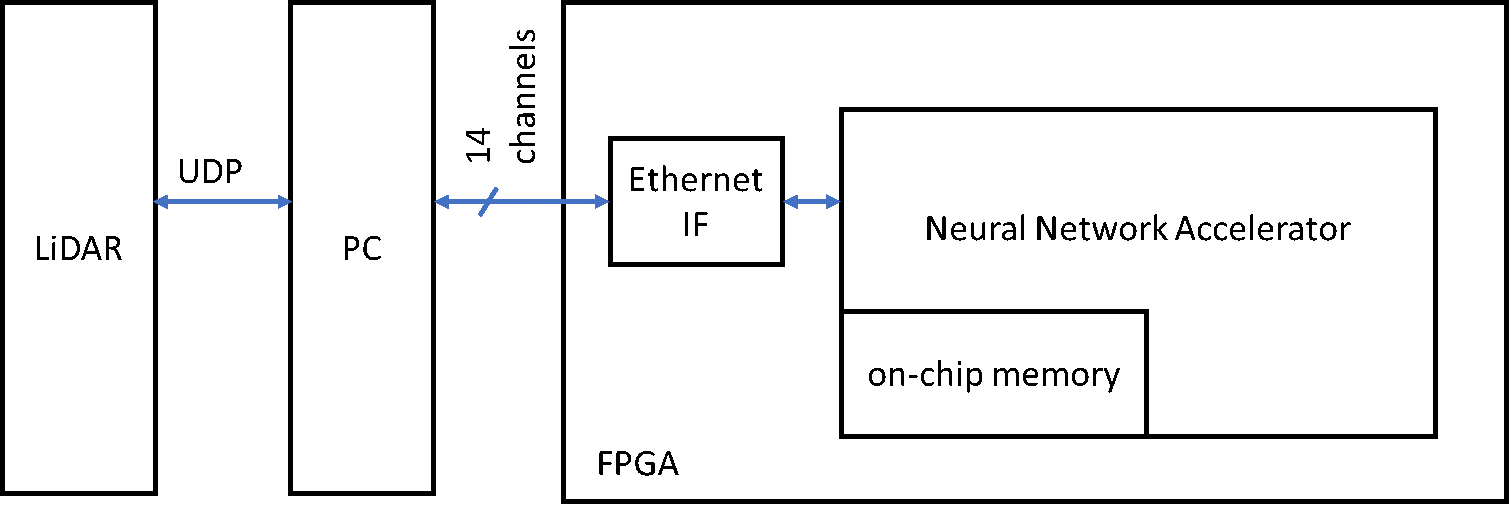}}

\textcolor{black}{\caption{The overall system architecture with a LiDAR and FPGA accelerator\label{fig:Test-system-architecture}}
}
\end{figure}

\textcolor{black}{The target hardware platform is Xilinx UltraScale
XCKU115 FPGA. An integrated test system is demonstrated in Figure
\ref{fig:Test-system-architecture}.}

\textcolor{black}{The LiDAR frames are transmitted into PC at 10 Hz
via UDP protocol. For each LiDAR frame, the PC pre-processes it and
sends an 18-bit feature map to the ChipNet neural network in the FPGA.
The feature map size is $64\times180\times14$. The parameters are
ported using MATLAB HDL coder. System clock frequency is set to 350
MHz. Each convolution block takes about 12,512 clock cycles to generate
2 feature maps. The total processing time of this CNN architecture
is about 12.59 ms. Since normally LiDAR point cloud frame rate is
10 Hz, this FPGA implementation fulfills the requirement of real-time
LiDAR data processing. When running ChipNet in software on the Intel
Core i5-5200U CPU, the processing time is 549 ms; when running ChipNet
using the NVidia K20 GPU, the processing time is 162 ms. Thus, the
FPGA implementation gains $43\times$ speed up over CPU and $13\times$
speed up over GPU. }

\textcolor{black}{As mentioned earlier, there are few FPGA implementations
of LiDAR processing using CNN at this time, performance and efficiency
comparison with similar works on FPGAs is not available. }

\textcolor{black}{The resource usage of our proposed neural network
is listed in Table \ref{tab:resource}. The total power consumption
of this design is 12.594 W, estimated by from Xilinx Vivado 2017.2
power analyzer using post-implemtation simulation .saif (Switching
Activity Interchange Format) file. As illsutrated in Table \ref{tab:power},
it consists of dynamic power 9.747W and static power 2.848W. We notice
that most of the power is consumed by the on-chip memories since they
are always enabled except in idle mode. We emphasize that the proposed
FPGA solution takes only 11.8\% power consumption of an NVidia K20
GPU, which is 107 W. }

\textcolor{black}{}

\textcolor{black}{}
\begin{table}[H]
\textcolor{black}{\caption{Resource usage on the FPGA implementation of ChipNet\label{tab:resource}}
}
\centering{}\textcolor{black}{}%
\begin{tabular}{|c|c|c|c|}
\hline 
\textcolor{black}{FPGA Resource} &
\textcolor{black}{Used} &
\textcolor{black}{Available} &
\textcolor{black}{Utilization}\tabularnewline
\hline 
\hline 
\textcolor{black}{Slice Registers} &
\textcolor{black}{33530} &
\textcolor{black}{1326720} &
\textcolor{black}{2.53\%}\tabularnewline
\hline 
\textcolor{black}{Slice LUTs} &
\textcolor{black}{38082} &
\textcolor{black}{663360} &
\textcolor{black}{5.74\%}\tabularnewline
\hline 
\textcolor{black}{Block RAMs} &
\textcolor{black}{1543} &
\textcolor{black}{2160} &
\textcolor{black}{71.44\%}\tabularnewline
\hline 
\textcolor{black}{DSPs} &
\textcolor{black}{3072} &
\textcolor{black}{5520} &
\textcolor{black}{55.65\%}\tabularnewline
\hline 
\end{tabular}
\end{table}

\textcolor{black}{}
\begin{table}[H]
\textcolor{black}{\caption{\textcolor{black}{Power estimation of FPGA design\label{tab:power}}}
}
\begin{centering}
\textcolor{black}{}%
\begin{tabular}{|c|c|c|}
\hline 
\textcolor{black}{Power Type} &
\textcolor{black}{Item} &
\textcolor{black}{Power Consumed}\tabularnewline
\hline 
\hline 
\multirow{4}{*}{\textcolor{black}{Dynamic}} &
\textcolor{black}{Logic} &
\textcolor{black}{0.212 W}\tabularnewline
\cline{2-3} \cline{3-3} 
 & \textcolor{black}{BRAM} &
\textcolor{black}{8.609 W}\tabularnewline
\cline{2-3} \cline{3-3} 
 & \textcolor{black}{DSP} &
\textcolor{black}{0.006 W}\tabularnewline
\cline{2-3} \cline{3-3} 
 & \textcolor{black}{MMCM} &
\textcolor{black}{0.26 W}\tabularnewline
\hline 
\textcolor{black}{Satistic } &
\multicolumn{2}{c|}{\textcolor{black}{2.848 W}}\tabularnewline
\hline 
\end{tabular}
\par\end{centering}
\end{table}

\section{\textcolor{black}{Conclusions\label{sec:Conclusions-and-future}}}

\textcolor{black}{In this paper, the problem of drivable region segmentation
is framed as a semantic segmentation task by processing real-time
LiDAR data using a convolutional neural network on an FPGA. The LiDAR
data is organized in spherical view and sampled to a dense input tensor
during pre-processing. An efficient and extendable CNN architecture
namely ChipNet is proposed as the main processor. A reusable and efficient
3D convolution block is designed for FPGA implementation. The proposed
approach is trained using Ford dataset and the KITTI benchmarks. Evaluations
show the proposed LiDAR processing algorithm can achieve state-of-art
performance in accuracy and also real-time processing in speed on
the FPGA. However, the FPGA implementation still consumes a large
amount of on-chip memory. For future work, we will consider recurrent
neural network for spatial-sequence decoding that may reduce the on-chip
memory usage. We also notice during benchmark evaluation that sidewalk
and railway are the main causes of false positives. Sensor fusion
of LiDAR and camera data will be considered to further improve the
accuracy.}

\textcolor{black}{\bibliographystyle{plain}
\bibliography{23_home_yecheng_TCAS1-2018_reference}
}
\begin{IEEEbiography}[{{{\includegraphics[clip,width=1in,height=1.25in]{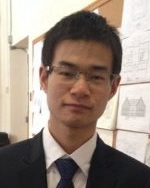}}}}]{Yecheng Lyu}
\textcolor{black}{(S'17)received his B.S. degree from Wuhan University,
China in 2012 and M.S. degree from Worcester Polytechnic Institute,
USA in 2015 where he is currently a Ph.D student working on autonomous
vehicles. His current research interest is sensor fusion, autonomous
vehicle perception and deep learning.}
\end{IEEEbiography}

\textcolor{black}{}

\begin{IEEEbiography}[{{{\includegraphics[clip,width=1in,height=1.25in]{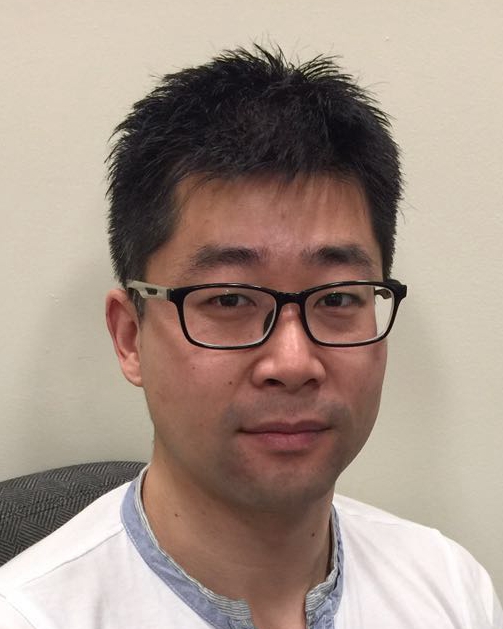}}}}]{Lin Bai}
\textcolor{black}{(S'18) received the B.S degree in integrated circuits
design and integrated system from University of Electronic Science
and Technology of China in 2009 and M.S. degree in electrical engineering
and information technology from Swiss Federal Institute of Technology
Zurich in 2012. Previously he was an FPGA engineer in industry. He
is currently working towards the Ph.D. degree at Worcester Polytechnic
Institute, USA. His current research interest is hardware acceleration
of deep learning algorithms on FPGA and ASIC. }
\end{IEEEbiography}

\textcolor{black}{}

\begin{IEEEbiography}[{{{\includegraphics[clip,width=1in,height=1.25in]{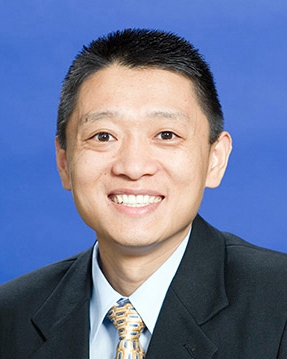}}}}]{Xinming Huang}
\textcolor{black}{(M\textquoteright 01\textendash SM\textquoteright 09)
received the Ph.D. degree in electrical engineering from Virginia
Tech in 2001. Since 2006, he has been a faculty in the Department
of Electrical and Computer Engineering at Worcester Polytechnic Institute
(WPI), where he is currently a chair professor. Previously he was
a Member of Technical Staffs with the Bell Labs of Lucent Technologies.
His main research interests are in the areas of circuits and systems,
with emphasis on autonomous vehicles, deep learning, IoT and wireless
communications. }
\end{IEEEbiography}

\end{document}